\documentclass[aps,pre,showpacs,showkeys,twocolumn]{revtex4}

\usepackage{graphicx}
\usepackage{latexsym}

\begin{document}

\title{Numerical studies on chaoticity of a classical hard-wall billiard 
	with openings}

\author{Suhan Ree}
\affiliation{Department of Industrial Information,
	Kongju National University, Yesan-Up, Yesan-Gun, 
	Chungnam, 340-802, South Korea}
\email{suhan@kongju.ac.kr}

\date{\today}

\begin{abstract}
Using fractal analysis, we investigate how the size of openings 
affects the chaotic behavior of
a classical closed billiard when two
openings are made on the boundary of the billiard.
This kind of open billiards retains chaotic properties of 
original closed billiards when openings are small compared to the size
of the billiard.
We calculate the fractal dimension using an 
one-dimensional subset of all possible
initial conditions that will produce
trajectories of a particle injected from an opening
for the billiard,
and then observe how the opening size 
can change the classical chaotic properties of this open billiard.
\end{abstract}

\pacs{05.45.Df, 05.45.Pq, 73.23.Ad}
\keywords{Chaos, Billiard, Classical scattering, Fractal}

\maketitle

Two-dimensional (2D) hard-wall billiard systems have been a popular subject
for studying the dynamics of chaotic systems\cite{reichl}. 
Theoretically, 2D billiard systems can be used to study manifestations of classical 
chaos in semiclassical and quantum mechanics.
The classical dynamics of a billiard system shows three distinct types of behavior:
the system is either integrable (\emph{regular} behavior) or non-integrable 
(either \emph{soft chaos}, characterized by mixed phase spaces that have 
both regular and chaotic regions, or \emph{hard chaos}, characterized by 
ergodicity and mixing)\cite{gutzwiller}. 
In experiments, on the other hand, the billiard systems can be used as models
to explain fluctuating behavior of magnetoconductance through
2D semiconductor heterostructures in the ballistic regime\cite{markus,chang}.
In such an experiment, the structure have the shape of a 2D billiard, of which the dynamic
properties of the billiard are already known (for example, the circle as
a regular system, or the stadium as a system with hard chaos).
To measure the conductance through the structure, leads 
(from here on, we will call them \emph{openings}) are attached to the structure;
hence the whole structure becomes an open billiard.
One usually studies transmission properties through these open billiards,
and links results to the classical chaotic properties of original
closed billiards, because the properties of the closed billiard
are still observed indirectly when the size of the openings is small 
compared to the size of the billiard
(see Refs.~\cite{baranger,ree2,fuchss}).
Therefore, to better relate the open billiard to the original closed billiard,
quantitative analysis on effects of openings in classical dynamics
is necessary.

In this paper, we numerically calculate a quantity that represents the chaoticity
of classical open billiards, and see how the size of the openings affects the overall 
classical dynamics. 
In Ref.~\cite{bleher}, the fractal dimension was obtained by looking at 
a two-dimensional set of initial conditions that will produce
trajectories injected from an opening of the Sinai billiard with two openings.
[A set of initial conditions in four-dimensional phase space can
be reduced to a two-dimensional set, because (1) energy of the particle
does not affect the trajectory of a particle, and (2) initial
locations of the particle can be restricted to an one-dimensional (1D) curve.]
Here, however, we will use
an 1D subset of all possible initial 
conditions for calculations of the fractal dimension to make 
numerical calculations simple.
The incident angle of the particle injected from the center of
one opening represents an 1D subset of 
all possible initial conditions,
which give rise to some possible trajectories of injected particles.
In Ref.~\cite{ree3}, graphs of the ``exit opening'' (an opening from which 
the particle exits) vs the incident angle from the center of an opening
were used.
There are ``geometrical channels''\cite{roukes,luna} inside a range of
incident angles.
When there are two openings, we can define
\emph{transmission windows} that represent initial angles for transmitting particles.
The boundaries of these windows will form a fractal\cite{bleher,ree3}.
In this paper, instead of calculating the fractal dimensions of boundaries of 
these transmission windows, 
we use the graph of the number of collisions vs the incident angle, and calculate the 
fractal dimension of the set of singularities, at which the number 
of collisions changes abruptly,
inside the set of all possible incident angles (see Ref.~\cite{eckhardt}).
The fractal dimensions from these two graphs will be eventually the same.

\begin{figure}
\includegraphics[scale=0.5]{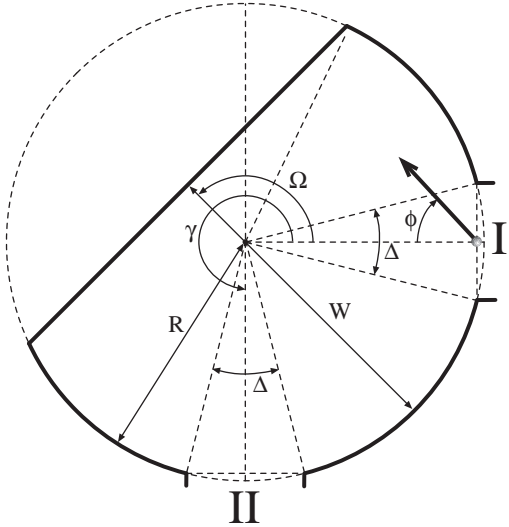}
\caption{\label{geometry}
The geometry of the cut-circle billiard. The
billiard has two openings, I and II. 
The size of the cut is given by the width $W$ (we define $w$ as $W/R$
to represent the relative size of the cut). 
The size of two openings is $\Delta$, and the position of 
the cut and the opening II, $\Omega$ and $\gamma$,
are measured from the position of the opening I. 
A particle is injected 
from the center of the opening I with an angle $\phi$.}
\end{figure}

The billiard used here for numerical calculations
is the circular billiard with a straight cut (\emph{the cut-circle
billiard}) with two openings
(see Fig.\ \ref{geometry}).
There are five parameters: (1) the width $W$ measured in the
direction perpendicular to the cut, (2) the radius $R$, (3) the angular
width $\Delta$ of the openings, which represents the opening size, 
(4) the orientation angle
${\Omega}$ of the cut relative
to the first opening, and (5) the position of the second 
opening relative to the
first opening as measured by the angle $\gamma$. 
We scale the width $W$ by $R$ so
$w\equiv W/R$, and thereby reduce the number of independent 
parameters to four:
$w$, $\Delta$, $\Omega$, and $\gamma$. For all subsequent 
discussions, we set
$\Omega=135^\circ$ and $\gamma=270^\circ$.
For these values, $w$ cannot be less than $[1-\cos(45^\circ)]\simeq0.293$, and 
$\Delta$ has an upper limit $\Delta_{max}$,
\begin{equation}
\label{deltamax}
\Delta_{max} =\left\{
	\begin{array}{ll}
		2\;[\cos^{-1}(1-w)-\pi/4]~ &\rm{when}~0.293<w<1\\
		\pi/2 			& \rm{when}~1\le w \le 2
	\end{array}
\right. ,
\end{equation}
in radian.
When the billiard is closed ($\Delta=0$), it has been proven
that the phase 
space is mixed (soft chaos) when $0<w<1$, and that the phase space is fully chaotic
(hard chaos) when $1<w<2$\cite{bunimovich}.  
The system is integrable when $w=1$ and 2.
This billiard has a property of showing all three types of dynamic behaviors
just by changing a parameter $w$, and this is the reason why the cut-circle billiard 
is a good example for analysis like this.

\begin{figure}
\includegraphics[scale=0.45]{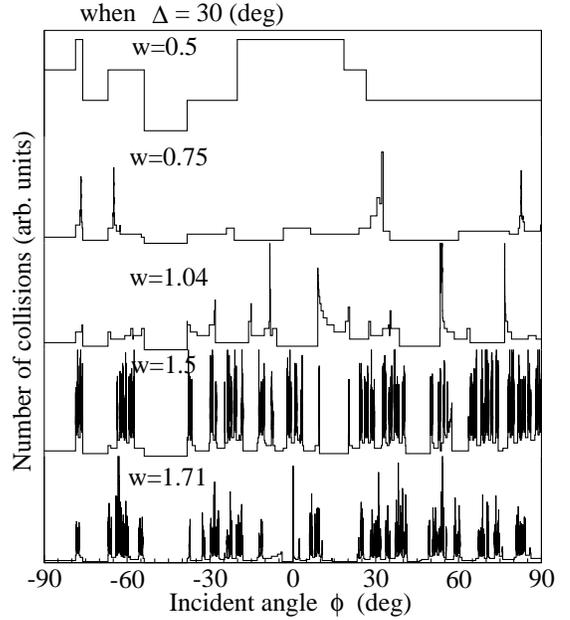}
\caption{\label{ncols}
Graphs of the number of collisions vs the incident angle  $\phi$.
Here the number of collisions is scaled for each $w$ to fit in
graphs with the same sizes, because
we are only interested in positions of singularities.
Five cases with different cut sizes are shown : $w=0.5$, 
$w=0.75$, $w=1.04$, $w=1.5$, and $w=1.71$.
}
\end{figure}

The particle is injected with an incident angle $\phi$ ($-\pi/2<\phi<\pi/2$).
One can easily calculate several quantities by numerically following these 
trajectories, but we focus only on
the number of collisions with the wall before the exit with respect to $\phi$.
In other words, we find a mapping from the incident
angle in opening I to the number of collisions before the exit.
In Fig.\ \ref{ncols}, we show how the number of collisions changes with $\phi$
for five different $w$ values ($w=0.5$, 
$w=0.75$, $w=1.04$, $w=1.5$, and $w=1.71$) when $\Delta=30^\circ$. 
In these calculations, there are finite number of singularities when $w=0.5$,
but, in other cases, there are regions where singular points are closely packed together
showing infinitely fine structures. 
We observe that there tends to be more singularities as $w$ increases. 
That is because (1) when $w$ is bigger than one, the 
original closed billiard shows hard chaos, which means global chaos, and 
(2) the opening size
is fixed here for all $w$'s even though the overall size of the billiard 
increases as $w$ increases, as a result, 
making the relative size of the openings smaller.

We calculate the fractal dimension $d_f$ for
sets of singularities for various cases,
using a simplified box-counting algorithm.
For $N_P$  uniformly distributed incident angles, we numerically find
the number of collisions for each incident angle representing an initial condition. 
For several $N_P$ values, we find the number of singularities $N_S$
(assuming that there only exists a singularity when the number of collisions changes 
for two successive incident angles).
Then the fractal dimension $d_f$ is defined by 
\begin{equation}
\label{df}
d_f\equiv\lim_{N_P\rightarrow\infty}\frac{\log_{10}{N_S}}{\log_{10}{N_P}},
\end{equation}
which is the slope of the graph of $\log_{10}{N_S}$ vs $\log_{10}{N_P}$.
We use the ordinary least-square fit 
to find the slope using points in the graph with limited $N_P$. 
Because the numerical accuracy for numbers is limited, $N_P$
have an upper limit.
Moreover the numerically given error of an initial condition will amplify as
the particle travels through the billiard and hits the boundary while traveling.
Here we use the double-precision numbers with 15 significant digits,
and then $N_P$ should be less than $10^{15}$, especially when the number 
of collisions is big.
In most cases in our example, $N_P$ up to $10^7$ were used
to obtain $d_f$'s with two significant digits. 
In some cases (for example,
when $w=1.71$ and $\Delta>59^\circ$), however, $N_P$'s up to $10^8$ were
used to find $d_f$'s, 
because the slopes didn't converge sufficiently
when $N_P$ is less than $10^7$.

\begin{figure}
\includegraphics[scale=0.45]{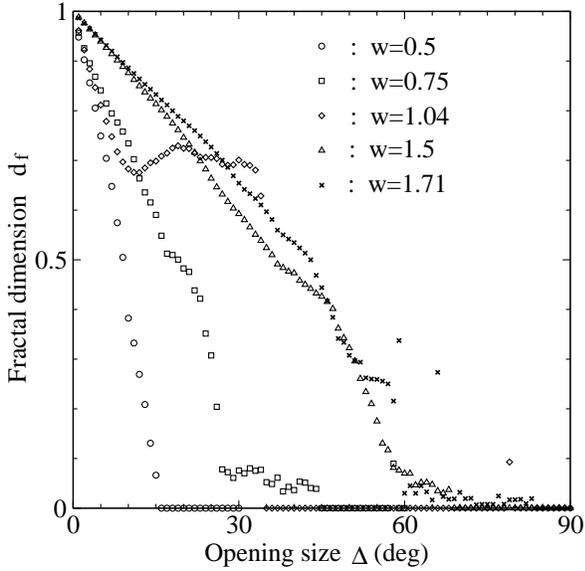}
\caption{\label{angles}
Graphs of the fractal dimension $d_f$ vs the opening size $\Delta$.
For each $w$ ($w=0.5$, $w=0.75$, $w=1.04$, $w=1.5$, and $w=1.71$), 
$\Delta$ is varied up to a maximum value $\Delta_{max}$. 
We can observe that $d_f$ does not always decreases monotonically
as the opening size $\Delta$ increases.
}
\end{figure}

In Fig.\ \ref{angles}, we fix the width $w$ at five different values used in Fig.\
\ref{ncols}, and vary the opening size $\Delta$ up to $\Delta_{max}$ for each $w$.
An interesting feature of these curves is that they are not monotonically decreasing
as $\Delta$ increases. Small peaks appear even after $d_f$ reaches zero 
at a certain $\Delta$ value. For example, the $d_f$-curve for $w=1.04$ reaches
zero near $\Delta\sim35^\circ$, but there is a small peak near $\Delta\sim79^\circ$.
The reason for this phenomenon is that there are more possible trajectories
when the opening is bigger. 
In our method for the circular billiard, 
the launching point of the particle 
gets closer to the center of the billiard as the opening 
size increases. 

\begin{figure}
\includegraphics[scale=0.45]{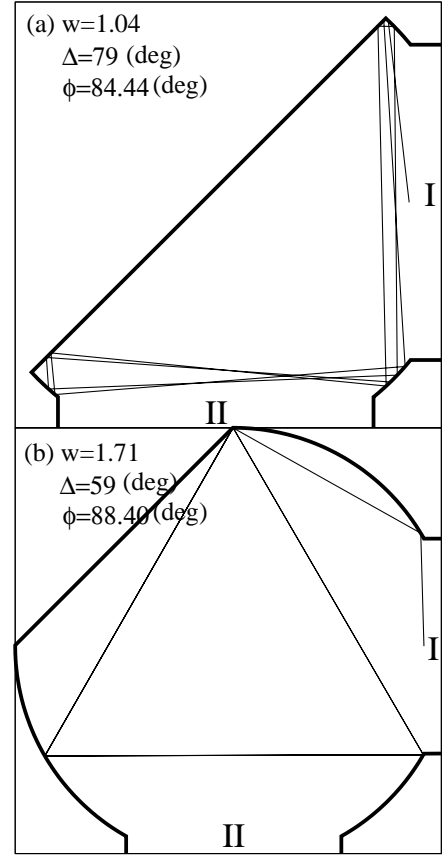}
\caption{\label{traj}
Two trajectories showing why the function of $d_f$ with respect to $\Delta$
does not always decreases monotonically. As the opening size $\Delta$ increases, new
initial conditions for trajectories, which can be very sensitive 
to the incident angle $\phi$, may appear. For example, 
(a) $w=1.04$, $\Delta=79^\circ$, and $\phi=84.44^\circ$. (b) $w=1.71$,
$\Delta=59^\circ$, and $\phi=88.40^\circ$. 
}
\end{figure}

In Fig.\ \ref{traj}, two trajectories that doesn't exist
when openings are smaller are shown.
Since these trajectories have the higher numbers of collisions and there 
is sensitive dependence on initial conditions near them, $d_f$ becomes non-zero.
In Fig.\ \ref{traj}(a), the billiard ($w=1.04$, $\Delta=79^\circ$) lost most
of the original shape of the closed billiard, but trajectories near 
the one shown ($\phi=84.44^\circ$) still make up an infinitely fine structure.
In Fig.\ \ref{traj}(b), the billiard ($w=1.71$, $\Delta=59^\circ$) has
a trajectory ($\phi=88.40^\circ$) 
that gets close to the 3-bounce closed orbit in the circular
billiard. (The closed orbits in the circular billiard are stable, but only
marginally\cite{berry,ree}.)
Even though only few bounces are shown, this trajectory actually bounces 
more than two thousand times as the triangular shape rotates clockwise little by little
for each rotation, before the particle finally exits through either one opening.
A high peak near $\phi=59^\circ$ for $w=1.71$ in Fig.\ \ref{angles} is due to 
trajectories near this one. 

\begin{figure}
\includegraphics[scale=0.45]{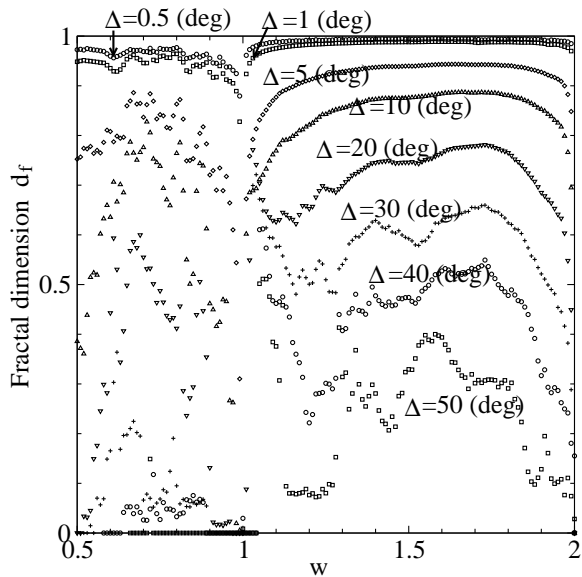}
\caption{\label{widths} 
Graphs of the fractal dimension $d_f$ vs $w$ for 8 different 
opening sizes, $\Delta=0.5^\circ$ ($\circ$), $\Delta=1^\circ$ ($\Box$), 
$\Delta=5^\circ$ ($\Diamond$),
$\Delta=10^\circ$ ($\bigtriangleup$), $\Delta=20^\circ$ ($\bigtriangledown$), 
$\Delta=30^\circ$ ($+$), 
$\Delta=40^\circ$ ($\circ$), and $\Delta=50^\circ$ ($\Box$). 
When $\Delta=5^\circ$ and $\Delta=10^\circ$, the behavior of the graph
is clearly distinct for two regions: 
hard chaos ($1<w<2$) and soft chaos ($0<w<1$). But when
$\Delta>20^\circ$, the distinction
between two regions disappears.
}
\end{figure}

In Fig.\ \ref{widths}, we calculate $d_f$'s as the width $w$ varies from 0.5
to 2 with the step size 0.01, 
for several opening sizes ($\Delta=0.5^\circ$, 
$\Delta=1^\circ$, $\Delta=5^\circ$, $\Delta=10^\circ$, 
$\Delta=20^\circ$, $\Delta=30^\circ$, $\Delta=40^\circ$, and
$\Delta=50^\circ$).
We can compare graphs in two different regions (found in 
the closed cut-circle billiard): $0<w<1$ (soft chaos)
and $1<w<2$  (hard chaos).
When $w=1$ and $w=2$ (integrable cases), $d_f$ is zero in our calculation.
As the opening size $\Delta$ approaches zero (see graphs for 
$\Delta=0.5^\circ$ and $\Delta=1^\circ$), we observe that the fractal 
dimension $d_f$ approaches one in both regions in our calculation.
When the opening size $\Delta$ is not big (see graphs for 
$\Delta=5^\circ$, $\Delta=10^\circ$, and $\Delta=20^\circ$),
we observe that behavior in two regions is clearly distinct.
In the region $0<w<1$, there are fluctuations, which comes from 
the mixed phase space structures
of the billiard, 
and in the region $1<w<2$,
graphs are smooth because the phase spaces of the billiard 
have no structure due to ergodicity.
On the other hand, when the opening size gets bigger 
(see graphs for $\Delta=30^\circ$,
$\Delta=40^\circ$, and $\Delta=50^\circ$), the  distinction between
two regions, observed in cases with smaller openings,  starts to disappear;
there are fluctuations in both regions. 
Ergodicity in closed cut-circle billiards with hard chaos is no longer 
an important factor in the dynamics of open billiards when openings are big,
as expected.

We have so far observed numerical results for the cut-circle billiard with
two openings. 
A method that calculates
the fractal dimension using an 1D subset of 
all possible initial conditions has been used to determine
effects of  openings on chaotic closed billiard quantitatively. 
From numerical results, we have found that the fractal dimension
as a function of the opening size is not a monotonically 
decreasing function,
because there are more trajectories, which can be sensitive to
initial conditions, for bigger openings.
These new trajectories can make up infinitely fine
structures, and subsequently can cause the fractal dimensions to increase 
with $\Delta$.
We also have found that, as the opening size gets bigger, the distinction
caused by soft chaos and hard chaos of the original  closed billiard
starts to disappear,
because the open billiard
that are constructed by introducing openings to the closed billiard
starts to lose the original shape of the closed billiard
as the size of openings gets bigger.

There are several points worth mentioning.
First, if we use the 2D full set of initial conditions to calculate
the fractal dimensions as was done in Ref.~\cite{bleher},
$d_f$ will be in the range of $1\le d_f \le 2$, but the overall 
behavior of $d_f$ is expected to be similar to that of the results obtained here.
Second, we can use a mapping from initial conditions to 
other quantities like dwell times (or travel distances) 
instead of the exit openings and
the number of collisions. Even though the dwell time
will be more expensive to calculate numerically, the set of 
singularities from the dwell time will be almost the same as those
from the number of collisions. 
Finally, it has been found to be important and necessary to understand 
relations between classical dynamics and quantum (or
semiclassical) dynamics of billiards\cite{blumel,ott2,jung},
because the ballistic scattering is usually realized in microscopic scales.
Then it will be interesting to ask how this kind of
fractal dimensions can be related to the quantum and semiclassical
dynamics of the same kinds of open billiards.

\begin{acknowledgments}
This work was supported by Kongju National University.
\end{acknowledgments}


\end{document}